\documentclass[sigconf]{acmart}
 \usepackage{multirow}
 \usepackage{graphicx}
\usepackage{subcaption}
 
\AtBeginDocument{%
  \providecommand\BibTeX{{%
    \normalfont B\kern-0.5em{\scshape i\kern-0.25em b}\kern-0.8em\TeX}}}

\setcopyright{acmcopyright}
\copyrightyear{2023}
\acmYear{2023}
\acmDOI{XXXXXXX.XXXXXXX}

\acmConference[Conference acronym 'XX]{Make sure to enter the correct
  conference title from your rights confirmation emai}{June 03--05,
2023}{Woodstock, NY}
\acmPrice{15.00}
\acmISBN{978-1-4503-XXXX-X/18/06}

 \usepackage{amsmath}

\begin{document}

\title{Personalized Category Frequency prediction for Buy It Again recommendations}

\author{Amit Pande, Kunal Ghosh, Rankyung Park}
\email{(amit.pande, kunal.ghosh, rankyung.park)@target.com}
\affiliation{%
  \institution{Data Sciences, Target Corporation}
  \streetaddress{7000 Target Pkwy}
  \city{Brooklyn Park}
  \state{Minnesota}
  \country{USA}
  \postcode{55445}
}

\renewcommand{\shortauthors}{Pande, et al.}

\begin{abstract}
   Buy It Again (BIA) recommendations are \textcolor{black}{crucial} to retailers to help improve user experience and site engagement by suggesting items that customers are likely to buy again based on their own repeat purchasing patterns. 
Most existing BIA studies analyze guests' personalized behaviour at item granularity. This finer level of granularity might be appropriate for small businesses or small datasets for search purposes. However, this approach can be infeasible for big retailers which have hundreds of millions of guests and tens of millions of items. For such data sets, it is more practical to have a coarse-grained model that captures customer behaviour at the item category level. In addition, customers commonly explore variants of items within the same categories, e.g., trying different brands or flavors of yogurt. A category-based model may be more appropriate in such scenarios. We propose a recommendation system called a \textit{hierarchical PCIC model} that consists of a \textit{personalized category model} (PC model) and a \textit{personalized item model within categories} (IC model). PC model generates a personalized list of categories that customers are likely to purchase again. IC model ranks items within categories that guests are likely to reconsume within a category.
The hierarchical PCIC model captures the general consumption rate of products using survival models. Trends in consumption are captured using time series models. Features derived from these models are used in training a category-grained neural network. 
 We compare PCIC to twelve existing baselines on four standard open datasets. PCIC improves NDCG up to 16\% while improving recall by around 2\%. We were able to scale and train (over 8 hours) PCIC on a large dataset of 100M guests and 3M items where repeat categories of a guest outnumber repeat items. PCIC was deployed and 
 A/B tested on the site of a major retailer, leading to significant gains in guest engagement.
\end{abstract}





\maketitle

\section{Introduction}

\begin{figure*}[htb]
    \centering
    \begin{subfigure}[b]{0.50\textwidth}
        \includegraphics[width=\textwidth, height=53mm]{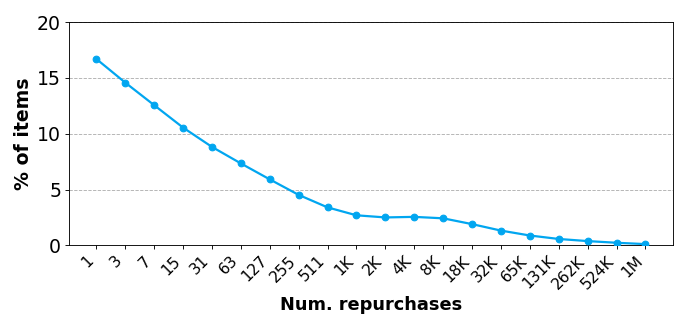}
        \vspace{-5mm}
        \caption{Percentage of customers repurchasing the same item}
        \label{fig:category_effectiveness_item}
    \end{subfigure}
    \begin{subfigure}[b]{0.49\textwidth}
        \includegraphics[width=\textwidth, height=53mm]{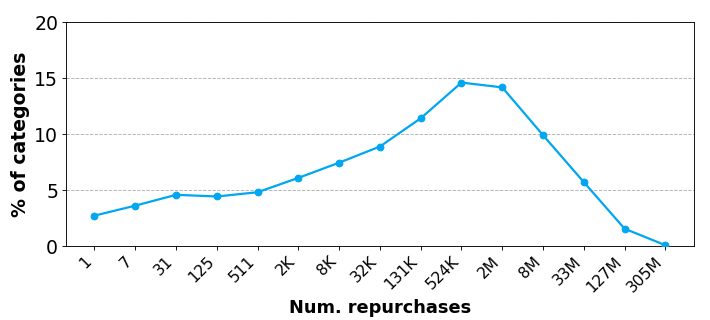}
        \vspace{-5mm}
        \caption{Percentage of customers repurchasing from the same category}
        \label{fig:category_effectiveness_category}
    \end{subfigure}

    \caption{Percentage of items and categories against number of repurchases in 1.5 years. {(a) Most items have small number of repurchasing transactions. (b) Most categories have large number of repurchasing transactions. Categories have more sufficient amount of data for modeling than items.}}
    \label{fig:category_effectiveness}

\end{figure*}
Note \footnote{The short version of this paper appears in RecSys 2023}

With the advent of e-commerce, recommendation systems have become a hot topic for research. Personalized recommendations are a key asset for successful apps or sites across a wide variety of industries including music or video streaming services, e-commerce platforms, gaming, finance, and banks. Digitization came late to the grocery shopping experience, as many people previously preferred to shop for groceries in person. However, digital grocery sales skyrocketed with the advent of Covid-19 as most shoppers switched to digital orders  backed by digital fulfillment, order-pickup, drive-up, or personal shopper~\cite{gomes2022evolution}. With this change in shoppers' behavior, a lot of attention went to both \textit{next basket recommendation} (NBR)~\cite{konstan1997grouplens,ren2019repeatnet,rendle2010factorizing,yu2016dynamic,ying2018sequential,hu2019sets2sets} that suggests items customers would like to purchase or consume next and to building personalized virtual aisles to aid the customer shopping experience. Effective personalized recommendations improve customer lifetime value (LTV) by increasing repeat purchases and by allowing customers to explore new relevant items.
 This brings a very good opportunity, especially for an omni-channel retailer, to design strategies which can keep them engaged by facilitating their shopping experience. Making the recurring purchases of customers quick is thus paramount to improve their shopping experience, and to free their time to purchase novel discretionary items.


Given a sequence of baskets that a customer has purchased or consumed in the past, the goal of a NBR system is to generate the next basket of items that the customer would like to purchase or consume next. Within a basket, items have no temporal order and are equally important. 
The NBR can be further divided into two similar but different problems. The first is repeat purchase recommendation, called the \textit{Buy It Again} (BIA) problem, where the goal is to recommend items that customers have already purchased and do so at times when the customers might be running out of the item(s). The second is adjacent inspiration recommendation, or the \textit{You might also like} problem, where the goal is to inspire customers to shop for items that may complement ones they have bought before or ones similar customers have purchased. 
Although many research papers on next basket recommendations lump the two subproblems together, most retailers implement them as entirely distinct products on their apps and webpages.


Existing work in BIA recommendations has focused on modeling item repurchase probabilities by using variants of recurrent neural networks or statistical models. Large retailers handle hundreds of millions of items and guests, but the majority of repurchase transactions are on a small subset of items and guests. This can lead to underfitting for item-grained models, as the data ends up being represented sparsely in a very high dimensional space. In the worst case, training itself may become infeasible due to computational resource limitations.


In this work, we emphasize the effectiveness of personalized category frequency modeling on BIA predictions. Customers will often explore variants of an item or new items within a category for reasons such as the desire to try different brands, the need to satisfy varying taste preferences in the customer's family, or the presence of discounts on alternative items. Category-based repurchase modeling can effectively capture higher abstraction information on these item repurchase dynamics. As shown in Figure~\ref{fig:category_effectiveness}, the percentage of items that have high numbers of repurchases is small (Figure~\ref{fig:category_effectiveness_item}), but most categories demonstrate high levels of repurchases (Figure~\ref{fig:category_effectiveness_category}). The discrepancy means that models geared toward category repurchases may be more effective at satisfying guest preferences. Furthermore, due to the aforementioned sparsity, it is far more difficult to train performant BIA recommendation models on item repurchases than it is on category repurchases. 

In this work, we emphasize the importance of both personalized product frequency as well as repeat purchase prediction models to make good Buy It Again predictions. More specifically, we observe that the product purchase frequency may be sparse in predicting customer repurchases and we discuss how using personalized category frequency can be a better choice. Customers often like to explore new items within a category. 



In this paper, we propose a 2-tier \textit{PCIC model} for BIA recommendations. The \textit{personalized category model} (PC model) predicts which categories customers will buy again on their next visit, and the \textit{personalized item within categories model} (IC model) provides personalized ranks of items in categories. Final BIA recommendations for individual customers are generated by combining both predictions. 
PC is a neural network that outputs category-level likelihoods for each customer. Input features to PC are generated by an ensemble of time-series machine learning algorithms that captures personalized consumption rates of each category and predicts when customers will buy items in each category. IC is a regression model that predicts category-agnostic item ranking. The outputs of the two models are combined to generate personalized BIA item recommendations for individual customers. 

We compare PCIC to twelve existing state of the art baseline algorithms on four standard open datasets. PCIC improves NDCG up to 16\% while improving recall by around 2\%. We were able to scale and train PCIC on a large dataset of 100M guests and 3M items where repeat categories of a guest outnumers repeat items. PCIC was deployed on an Apache Spark cluster, allowing us to train and score the model in around 8 hours. It was A/B tested on the site of a major retailer, leading to significant gains in guest engagement.

\vspace{2mm}
The main contributions of this work as summarized below: 
\begin{enumerate}
    \item {We propose a hierarchical PCIC model for Buy It Again recommendations which combines coarse prediction by a personalized category model (PC model) and finer-grained prediction by a personalized item within categories model (IC model). We show how the model supports our insights that customers tend to explore brands, sizes, flavors, etc. similar to a given item within a cateogy.}
    \item We demonstrate that the proposed PCIC model outperforms existing baselines of public datasets. We also show that PCIC scales to large datasets. 
    \item We deploy PCIC in a commercial setting to provide BIA recommendations for millions of customers. We demonstrate improved guest experience on the site as evidenced by multiple A/B tests. We discuss our experiences deploying and scaling PCIC. 
\end{enumerate}

\section{Literature Review}

One of the early reported work for Buy It Again recommendations came 
from Bhagat et al. ~\cite{bhagat2018buy} for Amazon shoppers' data 
back in 2018. In this work, the authors model the repeat consumption pattern of products using a modified Poisson-Gamma (mPG) model. The 
mPG model is built over a simpler PG model which assumes repeat-
purchase of a item at a customer level to be a possion process with a 
gamma prior for the purchase rate $\lambda$. They also provide two 
simple customer agnostic item level models viz. Repeat Customer 
Probability (RCP) and Aggregated Time Distribution (ATD) which works 
as a baseline for the mPG model in experiments. Another work was also reported by Dey et al. \cite{dey2016buy} in 2016, but this was more 
towards capturing repeat purchase behavior in longer time durations 
for e.g. several weeks to months. They have used PG model for capturing repeat purchase as base and then further used Dirichlet model to predict purchase probablities of items in a category.

Apart from the above work, we have been exploring other related works 
in the repeat purchase domain. While there were not so many, but still some notable works in the domain of customer purchase modeling has been done historically (starting from 60's era) where 
inspirations of modeling customer purchase events using statistical 
distributional assumptions can be taken. Once the mathematical 
expression of the unknown distributional parameters is rigorously derived, one can compute their estimates using data by calling simple 
math libraries / custom user defined functions etc. Several such 
works include, the Negative-Binomial distribution models (NBD) 
discussed in 
Enrehberg \cite{ehrenberg1959} and Grahn \cite{Grahn1969NBDMO}, the 
Erlang-2-Gamma model discussed by Chattfield and Goodhardt 
\cite{Chatfield1973ACP} etc. Later on, it was interesting to see works of Fader and Hardie on alternate versions of NBD model viz. 
Pareto-NBD, Beta-Geometric NBD \cite{faderhardielee2005}\cite{faderhardie2009} etc. While these 
approaches because of its strong foundations, may have influenced 
many later work based on statistical distributions (for e.g. 
\cite{bhagat2018buy}), but still these were mostly useful in solving some of the popular marketing problems (often referred as Marketing 
Science) like predicting shopping probabilities of a customer for the 
next n days tending to predict chances of their attrition, predicting 
expected customer basket size, predicting customer life-time value 
etc. The problems are mostly related to a customer's journey in a 
generic way and the solutions are often used to choose the right audience to whom retention policies needs to be deviced. When the notion of guest's category/item behaviors comes into the picture, 
(such as similar items, buy it again etc.), we should not be limited 
to such approaches. Rather using these approaches as signals and applying additional layers of learning with some supervision (if possible) would intuitively be a positive step to take.

Several literature on recommender systems are available, which has abilities to recommend a customer or user's personal taste on products. One of the older notable ones is the Grouplens project 
\cite{konstan1997grouplens} by Konstan et al. on Usenet news data in the late 90's, which used User based kNN (userKNN) approach of collaborative filtering to recommend personalized articles. Later on, 
another notable interesting approach we came across in the NBR domain was called the Factorised Personalised Markov Chain (FPMC) 
\cite{rendle2010factorizing} by Rendle et al. in 2010. This work uses 
a combination of two popular approaches to solve an NBR problem viz. 
Matrix Factorization (MF) which captures user's taste by factorizing 
observed user-item matrix and Marokov Chains (MC) which captures the 
sequential behavior of a user using transition graphs to predict the next action. Other similar works include, one by He et al. on 
sequential recommendation algorithms \cite{he2016fusing} in 2016 and another \cite{he2018translation} in 2018 which builds on the approach of \cite{rendle2010factorizing}. Another approach of using temporal 
dynamics on recommender algorithms was taken by Koren \cite{koren2009collaborative} in 2009, which is worth mentioning in this context. Our work certainly believes that temporal signals are important, but we have taken a different approach unlike integrating 
it directly with state-of-the-art recommender algorithms (viz. MF or 
MC) as done by \cite{rendle2010factorizing}, \cite{he2016fusing}, 
\cite{he2018translation} or \cite{koren2009collaborative}. We have 
considered or modeled it as separate signal and apply supervised learning on top it to cater to our problem.

More recently, with the popularity of neural network based applications, many other parallel and subsequent works have used a 
Recurrent Neural Network or LSTM or Transformer to more effectively capture the repeat purchase pattern. A more recent work by Hu et al. called Sets2Sets ~\cite{hu2019sets2sets} has the encoder which maps 
the set of elements from each previous time step onto a vector, while the decoder, uses a set-based attention mechanism to decode the set 
of elements from each subsequent time step from the vectors. This 
approach outperforms several state-of-the-art methods. Another work 
done by Hu et al. called TIFUKNN ~\cite{hu2020modeling} in 2020, propses a simpler method which outperforms even the RNN based approaches when it comes to NBR. It claims that personalized item frequency (PIF) provides critical signals for NBR, but existing methods including the RNNs fail to capture it. Their solution is an item frequency-based kNN method. It is to be noted that we also 
implement inter-category product ranking where item-frequency is a key signal, but our implementation is dependent on features derived from guest purchases while TIFUKNN depends on insights from similar guests using k-Nearest Neighbors. Another RNN approach developed by 
Yu et al. called DREAM ~\cite{yu2016dynamic} in 2016, where the input layer consists of multiple basket representations followed by a 
pooling operation on items in them to obtain a representation of the basket . Dynamic representation of the customer is obtained in the 
hidden layer and the output layer displays the customer's scores 
towards all items. The approach of Ying et al. called SHAN 
~\cite{ying2018sequential}, conststs of 2 stage attention layers 
called sequential hierarchical attention layers. The first layer 
captures customer's long-term behavior, followed by a second layer which is a composition of long and short term behavior. Finally we explore the approach of Ren et al. called RepeatNet 
~\cite{ren2019repeatnet} developed in 2019. They capture repeat 
consumption by incorporating a unique repeat-explore mechanism in 
RNN, which consists of encoder and 2 decoders to learn the recommendation probability for each item in the two modes viz. repeat and explore.  

There has been some work on hazard based approach by Kapoor et al. 
~\cite{kapoor2014hazard} in 2014, to predict customer's return time. 
They proposed framework to evaluate factors that influence customer 
return for web services, using the Cox's proportional hazard model 
~\cite{Cox1972Hazard}. This model can include several covariates. 
Compared to baseline regression and classification methods, the 
hazard-based model performs better in predicting user return time and 
categorizing users by their predicted return time. On top of this 
work, they also created a semi-Markov model ~\cite{Kapoor2015JustIT} 
that predicts when users will return to familiar items. The model takes into account latent psychological factors such as sensitization and boredom that occur when the same items are repeatedly consumed.

While we noted learnings from the existent research that has been done in the NBR domain, but as per the best of our knowledge our approach has its uniqueness and while compared against many of the above solutions as baselines, we saw promising results.
Our approach captures the importance of sequence models by considering time-series as a feature. 
It also accepts the success of hazard based approach and considers it to be an integral component of the solution.
Also, it takes care of PIF to generate recommendations at category to item level - which has been a concern for traditional RNNs. 
On top of it, it has capability to capture complex (non-linear) relationships amongst the all signals through a simple usage of FC neural network. 
    
\begin{table*}[htb]
\begin{tabular}{|c|ccccc|}
\hline
\hline

          & Num Items & Num Users  & Basket Size & Baskets/ User & Items/ user \\\hline

tafeng    & 12062     & 13949      & 6.27        & 5.69          & 6.397       \\\hline

dunhumby  & 4997      & 36241      & 7.33        & 7.99          & 22.56       \\\hline

shoppers  & 7907      & 10000      & 8.71        & 56.85         & 24.934      \\\hline

instacart & 8000      & 19935      & 8.97        & 7.97          & 33.271      \\\hline

Internal  & $\sim$3M  & $\sim$100M & $\sim$10    & $\sim$25      & $\sim$200  \\\hline
\hline

\end{tabular}
\caption{Some characteristics of datasets considered for evaluation}
\label{tab:datasets}

\end{table*}

\section{Model}

\subsection{Category level repurchase modeling}
We use category level features to predict the customers' likelihood to repurchase items. Each customer has their own features crafted by their purchase history, and the last \textit{m} days of customer purchase data is used to generate labels to train a category level model. All purchase history before this \textit{m} days is used to generate the features. Any category in which customers repurchased an item in this time period is considered label 1 while the other categories are assigned label 0. The main features considered to train the model are enumerated in subsequent subsections. The purchase history of a customer before this time frame is used to obtain features. 

\subsubsection{Survival Analysis}

Survival analysis focuses on the expected duration of time until occurrence of an event of interest. 
It differs from traditional regression by the fact that parts of the training data can only be partially observed, which is stated as being censored. 
For these censored observations, we only know that the event time is greater than the time at the point of censoring. In the retail scenario, we consider the purchase of an item within a category as an event. For each category, repeat purchase data can then be used to construct a life table across customers for each category, which will allow us to predict repeat purchase risk as a function of time. A life table summarizes the events and censored cases across time. At time 0, all observations (reference purchases) are still at risk, which meants that they have not yet repeated the purchase (event) or been censored. As events and censored cases occur, observations fall out of the risk set.

Repeat purchase data can be used to compute a few useful features:

1. \texttt{hazard} (eq.~\ref{eq:hazard}) is the probability of event occurring at kth day, conditional on the event not occurring before day k. It denotes an approximate probability that an event (repurchase) occurs in a given time interval, under the condition that an user would remain event-free up to that time (no purchase).
\begin{equation}
\texttt{hazard}_k = \texttt{n\_event}_k / \texttt{n\_risk}_k
\label{eq:hazard}
\end{equation}

2. \texttt{cum\_hazard} (eq.~\ref{eq:cum_hazard}) is cumulative sum of hazard over time. 
 
\begin{equation}
\texttt{cum\_hazard}_k = \sum_{kk = 0}^k \texttt{hazard}_{kk}
\label{eq:cum_hazard}
\end{equation}

3. \texttt{survival} (eq.~\ref{eq:survival}) is probability of the event occurring after day k or equivalently, the proportion that have not yet experienced the event by time t. 
 
\begin{equation}
\texttt{survival}_k = \exp(-1*\texttt{cum\_hazard}_k) 
\label{eq:survival}
\end{equation}

4. \texttt{cum\_survival} (eq.~\ref{eq:cum_survival}) as probability of event occuring in $\pm 3$ days to today. We additionally define this feature since many grocery customers shop once a week.
 
\begin{equation}
\texttt{cum\_survival}_k = \texttt{survival}_{k+3} - \texttt{survival}_{k-3}
\label{eq:cum_survival}
\end{equation}

5. \texttt{normalized\_risk} (eq.~\ref{eq:norm_risk}) is defined as risk associated with the user category today as a fraction of risk on the day of purchase.
\begin{equation}
\texttt{norm\_risk}_k = \texttt{n\_risk}_{k} / \texttt{n\_risk}_{0}
\label{eq:norm_risk}
\end{equation}

6. \texttt{normalized\_event} (eq.~\ref{eq:norm_event})  is defined as the event probability on the given day normalized by event plus censor population.
\begin{equation}
\texttt{norm\_event}_k = \texttt{n\_event}_{k} / \texttt{n\_event\_\&\_censor}_{k}
\label{eq:norm_event}
\end{equation}


Building this model gives a population level overview of the item repurchase rate. For example, we observe that people mostly repurchase bananas every 7 days and cleaning supplies every 21 days, so the hazard function is maximized at that time duration between purchases. Based on the last date of purchase of each item by the customer, we can use survival analysis to predict the date of repurchase or the probability of repurchase after n days. 

\subsubsection{ARIMA models}
Autoregressive Integrated Moving Average or ARIMA models are useful for short term forecasts on non-stationary time series problem. For each customer and category, we try to characterize their purchase pattern using ARIMA and predict the next day of purchase. ARIMA models have three parameters (\textit{p, d, q}) where \textit{p} is the order of the autoregressive model, \textit{d} is the degree of differencing, and \textit{q} is the order of the moving-average model. We build one ARIMA model that observes the past dates of purchases within a category to predict the next one and a second model to consider the quantity of item purchased and predict the current rate of consumption by the customer (say X uses 2 oz of shampoo daily). This is then used to predict the date when the customer will likely run out of the item. For each customer-category pair, we train these models and use their forecasts $\texttt{ARIMA(date)}$ and $\texttt{ARIMA(rate)}$ as features.

\subsubsection{Other features}
We consider three more behavioral category level features: $\texttt{NumPurchases}$ - Number of times a given customer has purchased from the category, $\texttt{tripsSinceLastPurchased}$ - the number of purchases in other categories customer has made since purchasing in this category, $\texttt{daysSincelastPurchased}$ - the time difference between today and last date the customer made a purchase in this category. 

\subsubsection{Model training}
We take the past 1.5 years of user shopping data to train the model to ensure we capture a yearly cadence. The last \textit{m} days of data is held out to generate labels. For example - we may take Jan 2021- July 24 2022 dataset to generate features for all guests. For those guests who shopped during July 25 - 31 (m = 7), we generate labels 0 and 1 for categories not shopped and shopped respectively. The 6 features from survival model, 2 predictions from two ARIMA models and the 3 other features mentioned earlier are generated for each user and category pair. 

We trained a 2 layer neural network on the category level guest purchase dataset. We wanted to keep it light because the number of input features is small (11), and we wanted it to scale well for the large number of users. 
The most performant neural net was composed of 2 fully connected layers (10 and 5 neurons) with sigmoid activations. The output layer is run through a softmax and the logistic loss function is used for optimization. 

\subsection{Inter-category Product Ranking}
In general, we observed that a customer is most likely to repurchase their most frequently or most recently bought items. The two main features used to rank products within a category are frequency (Freq) and recency (Rec) of purchase. We wanted to combine them both to arrive at optimal ranks, however, recency is measured in days and frequency is a count. To come to a common ground, we convert both into ranks. Item Frequency Rank (IFR) and Item Recency Rank (IRR) are obtained by ranking the frequency counts and days (respectively) since the last purchase of an item (DaysSincePurchase).
$\texttt{IFR} = Rk(Freq), \texttt{IRR} = Rk(DaysSincePurchase)$. We combine the ranks using a weighted average, rank again, then divide the rank by number of times the item is bought ($NIB$). This insight was based on user feedback and will be discussed in later sections. The equation ~\ref{eq:IR} shows how final Item Rank (IR) is calculated.
\vspace{-1mm}
\begin{equation}
\texttt{IR} =  ceil(\frac{1}{NIB}\times Rk(\alpha\times IRR + \beta \times IFR ))
\label{eq:IR}
\end{equation}
where the parameters $\alpha$ and $\beta$ were obtained using exhaustive grid search in the range [0,1].

\subsection{Model output}
We combine the outputs of PC and IC models to get an aggregated single list of items for recommendations. Let $Rk_{PC}$ and $Rk_{IC}$ represent the PC rank for an item's category and IC rank of the item respectively. The PCIC model outputs in a round robin manner i.e.
$Rk = Rk(sortByAscending(Rk_{PC},Rk_{IC}$))

\begin{table*}[]
\begin{tabular}{|c|cccc|cccc|}
\hline
                                    & \multicolumn{4}{c|}{Recall @10}                                                                                                                                                              & \multicolumn{4}{c|}{NDCG @10}                                                                                                                                                                \\ \hline
Dataset                            & \multicolumn{1}{c|}{Valued Shopper}                & \multicolumn{1}{c|}{Instacart}                     & \multicolumn{1}{c|}{Dunhumby}                      & TaFeng                        & \multicolumn{1}{c|}{Valued Shopper}                & \multicolumn{1}{c|}{Instacart}                     & \multicolumn{1}{c|}{Dunhumby}                      & TaFeng                        \\ \hline
TopSell                             & \multicolumn{1}{c|}{0.0982}                        & \multicolumn{1}{c|}{0.0724}                        & \multicolumn{1}{c|}{0.0819}                        & 0.0773                        & \multicolumn{1}{c|}{0.0779}                        & \multicolumn{1}{c|}{0.0641}                        & \multicolumn{1}{c|}{0.0601}                        & 0.0519                        \\ \hline
FBought                             & \multicolumn{1}{c|}{\textit{0.2109}}               & \multicolumn{1}{c|}{\textit{0.3426}}               & \multicolumn{1}{c|}{\textit{0.1853}}                       & 0.0704                        & \multicolumn{1}{c|}{\textit{0.2128}}               & \multicolumn{1}{c|}{\textit{0.3618}}               & \multicolumn{1}{c|}{{0.1771}}               & 0.0766                        \\ \hline
userKNN                             & \multicolumn{1}{c|}{0.0988}                        & \multicolumn{1}{c|}{0.0720}                        & \multicolumn{1}{c|}{0.1135}                        & {0.1089}               & \multicolumn{1}{c|}{0.1415}                        & \multicolumn{1}{c|}{0.1020}                        & \multicolumn{1}{c|}{{0.1707}}               & {0.0832}               \\ \hline
RepeatNet                           & \multicolumn{1}{c|}{0.1031}                        & \multicolumn{1}{c|}{0.2107}                        & \multicolumn{1}{c|}{0.1324}                        & 0.0645                        & \multicolumn{1}{c|}{0.1439}                        & \multicolumn{1}{c|}{0.2285}                        & \multicolumn{1}{c|}{0.1545}                        & 0.0592                        \\ \hline
FPMC               & \multicolumn{1}{c|}{0.0951}                        & \multicolumn{1}{c|}{0.0763}                        & \multicolumn{1}{c|}{0.0919}                        & 0.0868                        & \multicolumn{1}{c|}{0.1188}                        & \multicolumn{1}{c|}{0.0946}                        & \multicolumn{1}{c|}{0.1025}                        & 0.0667                        \\ \hline
DREAM      & \multicolumn{1}{c|}{0.0991}                        & \multicolumn{1}{c|}{0.0866}                        & \multicolumn{1}{c|}{0.0915}                        & 0.0902                        & \multicolumn{1}{c|}{0.1231}                        & \multicolumn{1}{c|}{0.1063}                        & \multicolumn{1}{c|}{0.1009}                        & 0.0763                        \\ \hline
SHAN           & \multicolumn{1}{c|}{0.0847}                        & \multicolumn{1}{c|}{0.0902}                        & \multicolumn{1}{c|}{0.1007}                        & 0.0878                        & \multicolumn{1}{c|}{0.1032}                        & \multicolumn{1}{c|}{0.1152}                        & \multicolumn{1}{c|}{0.1149}                        & {0.0813}               \\ \hline
Sets2Sets       & \multicolumn{1}{c|}{0.1259}                        & \multicolumn{1}{c|}{\textit{0.3021}}               & \multicolumn{1}{c|}{\textit{}{0.2068}}               & 0.1190               & \multicolumn{1}{c|}{\textit{0.1626}}                        & \multicolumn{1}{c|}{{0.3487}}               & \multicolumn{1}{c|}{\textit{0.2134}}               & {0.0844}               \\ \hline
{ TIFUKNN (Next Basket Recs)} & \multicolumn{1}{c|}{{ \textbf{0.3578}}} & \multicolumn{1}{c|}{{ \textbf{0.3952}}} & \multicolumn{1}{c|}{{ \textbf{0.2087}}} & { 0.1301} & \multicolumn{1}{c|}{{ 0.3060}} & \multicolumn{1}{c|}{{ \textit{0.3825}}} & \multicolumn{1}{c|}{{ \textit{0.1983}}} & { 0.1011} \\ \hline
TIFUKNN(BIA items only)     & \multicolumn{1}{c|}{\textit{0.3500}}               & \multicolumn{1}{c|}{\textit{0.3700}}               & \multicolumn{1}{c|}{\textit{0.1940}}               & 0.0990                        & \multicolumn{1}{c|}{\textit{0.3000}}               & \multicolumn{1}{c|}{\textit{0.3800}}               & \multicolumn{1}{c|}{\textit{0.1860}}               & {0.0860}               \\ \hline
RCP                                 & \multicolumn{1}{c|}{0.0416}                        & \multicolumn{1}{c|}{0.1090}                        & \multicolumn{1}{c|}{0.0635}                        & \textbf{0.3860}               & \multicolumn{1}{c|}{0.0591}                        & \multicolumn{1}{c|}{0.1175}                        & \multicolumn{1}{c|}{0.0634}                        & \textbf{0.2363}               \\ \hline
ATD                                 & \multicolumn{1}{c|}{0.0350}                        & \multicolumn{1}{c|}{0.1600}                        & \multicolumn{1}{c|}{0.0468}                        & \textit{0.3100}               & \multicolumn{1}{c|}{0.0605}                        & \multicolumn{1}{c|}{0.1264}                        & \multicolumn{1}{c|}{0.0350}                        & \textit{0.2310}               \\ \hline
PG                                  & \multicolumn{1}{c|}{0.1694}                        & \multicolumn{1}{c|}{0.2375}                        & \multicolumn{1}{c|}{\textit{0.1332}}                        & \textit{0.3100}               & \multicolumn{1}{c|}{0.0684}                        & \multicolumn{1}{c|}{0.1331}                        & \multicolumn{1}{c|}{0.0351}                        & \textit{0.2336}               \\ \hline
MPG            & \multicolumn{1}{c|}{{0.1762}}                        & \multicolumn{1}{c|}{{0.2183}}                        & \multicolumn{1}{c|}{0.0820}                        & \textit{0.3200}               & \multicolumn{1}{c|}{0.0680}                        & \multicolumn{1}{c|}{0.1240}                        & \multicolumn{1}{c|}{0.0450}                        & \textit{0.1600}               \\ \hline
PCIC model                         & \multicolumn{1}{c|}{\textit{0.3528}}               & \multicolumn{1}{c|}{{0.2548}}               & \multicolumn{1}{c|}{{0.1540}}               & {0.1427}                        & \multicolumn{1}{c|}{\textbf{0.3531}}               & \multicolumn{1}{c|}{\textbf{0.5700}}               & \multicolumn{1}{c|}{\textbf{0.2321}}               & {0.1180 }                       \\ \hline
\end{tabular}
\vspace{3mm}
\caption{Performance comparison with existing baselines. The top performing algo in a dataset are in bold. The three runner ups are in italics.}
\label{tab:performance}
\vspace{-1mm}
\end{table*}

\section{Experiments}
In this section, we conduct experiments to answer the following questions:
Q1: What is the effectiveness of the proposed method? Does it outperform state-of-the-art NBR/ BIA methods?
Q2: How well does this method scale up to generate recommendations for millions of users?
Q3: How is model performance impacted by the input features?
Q4: How do training and testing date ranges change the performance of the model?

\subsection{Experimental Settings}
\subsubsection{Datasets}
We use four publicly available datasets shown in Table~\ref{tab:datasets} 
to compare the performance of the proposed method with existing methods in literature: ValuedShopper\footnote{
https://www.kaggle.com/c/acquire-valued-shoppers-challenge/overview}, Instacart\footnote{https://www.kaggle.com/c/instacart-market-basket-analysis}, Dunnhumby\footnote{https://www.dunnhumby.com/careers/engineering/sourcefiles}, and TaFeng\footnote{https://www.kaggle.com/chiranjivdas09/ta-feng-grocery-dataset}. We also evaluate using an internal dataset consisting of the sales history of users at a large retailer. There are around 100M users and 3M products in this dataset.

\subsubsection{Evaluation Protocol}
We use recall (@K) and NDCG (@K) metrics to evaluate and compare our methods. The first metric evaluates the fraction of ground truth items, which customers bought in last trip, that have been rightly ranked over top-K items in all testing sessions. NDCG is a ranking based measure which takes into account the order of purchased items in the recommendations and generates a score between 0 to 1. We use the past baskets of a given customer to predict their last basket. We consider 80\% of customers data to train the model and remaining to test using 5-fold cross validation. We reserve 10\% training data as a validation dataset for hyper-parameters tuning in all the methods.

\subsubsection{Baselines}

\begin{enumerate}
    \item TopSell: It uses the most frequent items
that are purchased by users as the recommendations to all users.
\item FBought: It uses the most frequent items
that are purchased by a user as the recommendationto him.

\item userKNN~\cite{konstan1997grouplens}: It uses classical collaborative filtering based on kNN.
All the items in the historical baskets of user are merged as a set of items. 

\item RepeatNet~\cite{ren2019repeatnet}: RNN-based model for session-based
recommendation which captures the repeated purchase behavior of users. It uses GRUs and Attention.
To apply this method, user baskets are translated to a sequence of items. 

\item FPMC~\cite{rendle2010factorizing}: Matrix Factorization
uses all data to learn the general taste of the user whereas
Markov Chains can capture sequence effects in time. FPMC combines the both for Next Basket Recommendation problem.

\item DREAM~\cite{yu2016dynamic}: 
Dynamic REcurrent bAsket Model (DREAM) learns a dynamic representation
of a user but also captures global sequential features among
baskets. 

\item SHAN~\cite{ying2018sequential}: A deep model based on hierarchical attention networks. It partitions the historical baskets into longterm and short-term parts to learn the long-term preference and short-term preference based on the corresponding items attentively. 

\item  Sets2Sets~\cite{hu2019sets2sets}: The state-of-the-art end-to-end method for following multiple baskets prediction based on RNN. Repeated
purchase pattern is also integrated into the method.

\item RCP~\cite{bhagat2018buy}: Repeat Customer Probability (RCP) finds repeat probably of an item \& repeat items based on that.

\item ATD~\cite{bhagat2018buy}: Aggregate Time Distribution Model fits a time distribution to model probablity distribution and time characteristics of repeat items. 

\item PG~\cite{bhagat2018buy}: Poisson Gamma distribution fitted to predictaggregate purchasing behavior. 

\item MPG~\cite{bhagat2018buy}: A modified PG distribution to make the results time dependent and intergate repeat customer probability.

\end{enumerate}

We use grid search to tune the hyper-parameters in compared methods. For userKNN, the
number of nearest neighbors is searched from range(100, 1300). For FPMC, the dimension of factor is searched from the set of values [16, 32, 64, 128]. For RepeatNet, DREAM, SHAN, and Sets2Sets, the embedding size is searched from the set of values [16, 32, 64, 128]. For PCIC model, ARIMA model was autofitted in range (3, 3, 0). 

\vspace{-2mm}
\subsection{Performance Comparson (Q1)}

Table~\ref{tab:performance} gives the performance comparison of PCIC model with existing baselines. Several observations can be made from the table. 

First, we observe that the PCIC model has highest recall and NDCG values in most cases on Valued Shopper, instacard and Dunhumby datasets. 
Surprisingly, RCP model performs well on tafeng dataset. 

TIFUKNN models also performs well. Since that model is built for next basket recommendation task, whose results are under dataset TIFUKNN(NBR) in Table~\ref{tab:performance}, we modified the code and ran it to run on the BIA task only (i.e. generate user embedding vectors, find scores from neighbor embeddings and then filter out recommendations which user has not purchased before), whose results are under dataset TIFUKNN(BIA). We see that this leads to a slight dip in its performance. Just like our model captures personalized category frequency, TIFUKNN model tries to explicitly capture personalized item frequency. TIFUKNN model uses nearest neighbor approach to collaborative filtering to learn repurchasing pattern from other users. In PCIC model, the survival analysis features use user repurchasing pattern at category level. 

Sets2Sets captures personalized item frequency explicitly but subsequently learns coeffients for RNN. RCP, ATD, PG and MPG models do not use personalized item frequency but they try to model repeat purchase pattern using a Poisson Gamma or modified Poisson Gamma distribution. Hence, we can see that these methods perform better than any existing methods which do not capture item or category frequency such as RepeatNet, userKNN, etc.

FBought is a pretty simple baseline in that it simply ranks the most frequently bought items of a user in that order. It surprisingly performs better than many baselines here. It is a simple to implement baseline and performs pretty well. 

We wanted to select the best baselines and compare performance on a much larger, real-world internal dataset. The challenges in scaling these models to score for large datasets is discussed next. 

\vspace{-2mm}
\subsection{Scaling up (Q2)}

We attempted to train the top performing models above on a much larger (100M user) data set. TIFUKNN uses a user embedding the size of the entire product catalog, which made it impossible to scale up to this data set. Hence, it was impossible to scale TIFUKNN to run on this large dataset. Similarly, Sets2Sets uses GRU layers with attention and training on this data set would have taken weeks. As a result, we subsampled the larger data set, creating a representative sample with 1M users. We compared TIFUKNN and Sets2Sets to PCIC using this subsampled data. We observed a 30-35\% reduction in NDCG and recall metrics in TIFUKNN and Sets2Sets against PCIC. As a result, we did not put effort into scaling either algorithm.


PCIC was implemented in a distributed hadoop cluster using Apache Spark and takes around 6-8 hours of time to train and test the model for 100M users. The main time-consuming part is to figure out ARIMA model hyper-parameters for each user-category pairs and to generate those features. FBought is straight forward to implement and takes few minutes of run time. We also implemented MPG model in distributed cluster using the maths described in the paper. Table~\ref{tab:internal} shows the performance comparison of FBought, MPG and PCIC models. Although PCIC performs well in terms of NDCG, the recall is slightly lower than MPG . Next, we calculated MPG parameters at category level instead of original item level and input it as part of features to PC. The performance of integrated PCIC(+MPG) outperforms both PCIC and MPG. 

\begin{table}[]
\begin{tabular}{|l|l|l|l|l|}
\hline
& Recall@3        & NDCG@3          & Recall@5        & NDCG@5          \\ \hline \hline
FBought & 0.2020          & 0.0832          & 0.0305          & 0.1212          \\ \hline
MPG     & \textbf{0.0307} & 0.1036          & \textbf{0.0433} & 0.1328          \\ \hline
PCIC    & 0.0267          & \textbf{0.1071} & 0.0377          & \textbf{0.1368} \\ \hline \hline
PCIC(+MPG)    & 0.0317          & \textbf{0.1091} & 0.0447          & \textbf{0.1408}\\ \hline
\end{tabular}
\vspace{2mm}
\caption{Performance comparison on internal dataset}
\label{tab:internal}
\vspace{-1mm}
\end{table}

\vspace{-1mm}
\subsection{Feature Importance (Q3)}

\begin{figure}[htb]
  \centering
  \includegraphics[width=\linewidth, height=40mm]{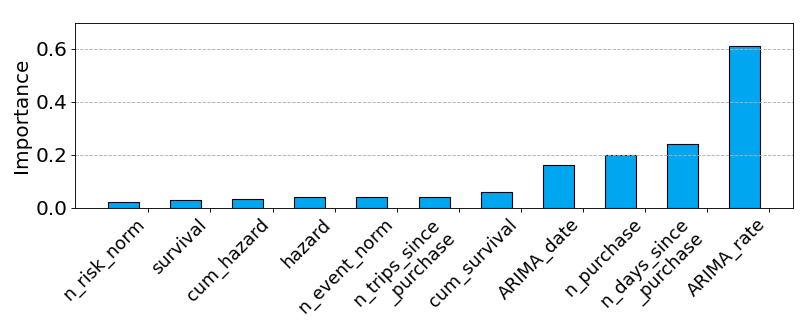}
    \vspace{-5mm}
  \caption{Relative importance of input features to PC model}
  \label{fig:fi}
  \vspace{-1mm}
\end{figure}

To obtain the feature importance, we replaced the original neural layer with a Gradient Boosting Tree clasifier. The values are plotted in figure~\ref{fig:fi}. We can observe that the ARIMA forecasts have a very high impact on the output of the model, particularly the model that tries to predict the next purchase based on rate of individual consumption of item by the user. The survival features have smaller impact on the prediction quality meaning other user's purchases play a small role in user's repurchase than his own characteristics. This can be one of the reason why approaches like itemKNN or TIFUKNN which focus on collaborative user behavior don't perform as well as PCIC. MPG does capture rate of consumption with a statistical model and it comes close to PCIC. The features such as number of days since past purchase and explicit category frequency (num purchase) also have high feature importance. if we were to collect the top 3 features, we can say that we can predict whether a user will purchase an item today based on how many times he has purchased before, how many days since his last purchase with us, how much did he purchase last time and how long will it last. 

\subsection{Impact of train and test data selection (Q4) }
We held out one week of the most recent customer purchases from this dataset for testing and used one year of purchases made prior to that week for training. A customer and their product purchase were considered as a repeat purchase in the test period only if the customer purchased a product in the training period (y years before the test period, y =1.5) and also purchased the same product sometime in the test period. The (user, category) pairs purchased in this duration are labeled 1 and the categories the user did not purchase in this duration was labeled as 0. 

As the pandemic caused increased adoption of the app and website, users started shopping online more frequently particularly. Based on the initial feedback, we observed that the BIA list was not updating particularly for the highly engaged users. We hypothezised that this can be because of the following reasons: (1) the model being trained on all users may not be able to exactly capture the signals and behavior of highly engaged user. (2) The labels are captured based on last 1 week of purchases. But highly engaged users shop much more often, hence their labels are not very accurate. 
We experimented with scoring the model daily on 1 day of user purchases. We also experimented on training the model only on the most engaged users, defined as users who have made purchases in more than 25 categories. 

Table~\ref{tab:testtrain} shows the improvement in NDCG metric for the PC model with the changes in test time frame and with training on only the most engaged users. Reducing the test time frame significantly improved the performance of the model. The most engaged users had a lower NDCG performance than all users when the test dates were 7 days. We also observed that training the model only on the most engaged users improves NDCG for all users too although it leads in savings on training time. The time taken to train the generate the features and train the model on all users is 2.5x the time taken for highly engaged users

\begin{table}[]
\begin{tabular}{|l|l|ll|}
\hline
\multicolumn{1}{|c|}{\multirow{2}{*}{Trained on}} & \multicolumn{1}{c|}{\multirow{2}{*}{Test Timeframe}} & \multicolumn{2}{l|}{NDCG (Test)}             \\ \cline{3-4} 
\multicolumn{1}{|c|}{}                            & \multicolumn{1}{c|}{}                                & \multicolumn{1}{l|}{Most Engaged}    & All \\ \hline
All                                               & 7 days                                               & \multicolumn{1}{l|}{0.2009} & 0.2325       \\ \hline
All                                               & 1 day                                                & \multicolumn{1}{l|}{0.3501} & 0.3583       \\ \hline
Most Engaged                                      & 1 day                                                & \multicolumn{1}{l|}{0.3602} & 0.3589       \\ \hline
\end{tabular}
\vspace{2mm}
\caption{Modifications in performance of PC model with changes in training data selection and testing timeframe.}
\label{tab:testtrain}
\vspace{-6mm}
\end{table}

\section{Deployment Journey}
In this section, we discuss several user-facing questions we addressed as well as our experience in deploying PCIC.

\subsection{Deployment and Online Experience}

While offline metrics are informative and help us build competent models, the true test of a recommendations model is online, where we can measure impact on user behavior. We deployed PCIC to a production environment where recommendations are generated daily in our compute cluster on an Apache Spark ecosystem and exported to the cloud for real-time serving. When a user visits the site, these recommendations are then served to them, filtered on the item availability based on inventory and available shipment options selected by the user.

\begin{table}[]
\begin{tabular}{|l|l|}
\hline
                                            & Lift (\%)                                               \\ \hline
CTR &   6 \\ \hline
Conversion &  8.5 \\ \hline
Units & 27.5 \\ \hline
\end{tabular}
\caption{Measuring impact of BIA against FBought.}
\label{tab:test_bia}
\vspace{-3mm}
\end{table}






\subsection{Human-in-the-loop feedback}
We first rolled out the results to a pool of internal team members for testing. This gave us some feedback as to having an exclusion list of some categories which users may not be very comfortable looking at, in their App (with friends and family or otherwise). Based on the feedback, we built an exclusion list of categories which are applied on top of recommendations as filters. 

Secondly, we found that users were sometimes recommended an item they'd recently purchased (e.g., a new flavor of yogurt) from a category where they repurchase, but not one they'd like to re-purchase. We used a two step approach filter out such items from recommendations. Apart from the category being a repurchase category, we tried to ensure that the item was bought by the guest at least twice in the past n months (n=6). This helps the customer to identify the items in buy it again list as an item they have repeat purchased. Second, we identified items with low repurchase rates (similar to repurchase rate threshold in RCP~\cite{bhagat2018buy}) and removed them. 

Several users also noted they typically buy more than one item from a specfic category (e.g., two or more flavors of yogurt) in a single trip. In the backend, we may have a ranked list of categories and ranked list of items within each category. Originally, the PCIC model would round robin among these lists to merge a new list which has first item in each category followed by second item on each category and so on till the list is finished. For a user who purchases more than one item per category every trip, this may be inconvenient. To resolve this, we calculate a variable $NIB$ which denotes the number of times the item was purchased by the user per trip. We tweaked the math used to combine the two lists by dividing item rank by NIB in \label{eq:IR} and then taking a ceil function to create new item ranks. 

\subsection{Metrics}
To quantify the impact of the proposed PCIC algorithm, we performed A/B tests against existing online baselines. Each test was run for more than two weeks and stopped after ensuring that the samples are statistically significant. The metrics considered for tests are defined as follows:
\begin{itemize}
\item CTR or Click Through Rate : Percentage of recommendation displays which were clicked by the guest.
\item Conversion Rate: Percentage of clicked recommendations which were purchased by the guest same day. 
\item Units: The total number of units purchased by the users who were part of the treatment.
\end{itemize}

\subsection{Testing against baseline}
When we introduced Buy It Again recommendation lists to the guest shopping experience, we A/B tested PCIC against a baseline of FBought. 
The results are given in Table~\ref{tab:test_bia}. We can see that there is significant lift across all three metrics - 6\% in CTR, 9\% in Conversion and 27\% in units purchased.

\subsection{Testing Buy it Again on web search}
We tested adding a Buy It Again recommendation list to the search results of all users. For this, we filtered the Buy It Again results using the search query context, so if someone searched for paper towel, the BIA recommendation list would be filtered to show only paper towels. As a sizable fraction of user searches do not pertain to items the user has already purchased, most of the time this recommendation list would not be shown to the guest. We found that the user interaction with this recommendation list was significantly higher than existing search results (by over 20\%). As we were testing a recommendation list against a non-existing one, we used visit level metrics to evaluate BIA. It was observed that the add-to-carts, average order values, and units per order went up by 0-2\%. We also observed that the guests were able to directly add the items to cart from the recommendation list and subsequently browsed fewer items despite having higher add to carts. These metrics consolidate our belief that showing buy it again items helps the guest in their shopping experience. 
\vspace{-0.1in}
\subsection{Building virtual aisles}
Research studies and our internal surveys indicate that online grocery shopping experience for users is significantly different from the typical user store experience~\cite{chintala2022browsing}. Shopping basket variety is significantly lower for online shopping trips, as measured by the number of unique categories and items purchased. Online grocery shopping environments may accelerate consumer inertia, leading to repurchase of essentials and reduction in purchase of fresh vegetables, impulse purchases such as candy or bakery desserts, and discretionary spending.

After identifying these opportunities in online shopping experience development, in 2021, we rolled out BIA to guests by filtering recommendations by categories (Milk, Yogurt, Beauty, etc) to create a virtual aisles experience for online users. We use the personalized list of categories for each guest using PC model. For each category, we present a list of recommended items from IC model to form a virtual aisle. In each aisle, we first showed the BIA items of the guest followed by other relevant items (generated using other algorithms and personalized to each guest, not discussed here for sake of brevity). 
We rolled out the experience directly to the users and report the lift in experience of guests who used the experience against those who didn't interact with it. 

 Users who interacted with these recommendations had a significant increase in units per order (25-50\%), and average order value (7-35\%). Since the buy it again essentials are lower ticket items, they have a smaller dollar impact in order value than units per order. We saw higher engagement of guests with virtual aisles experience for frequency cateogories in the App than in the site.

\section{Future Directions}

Buy It Again recommendations help users to quickly complete their shopping missions. Traditional approaches tend to model guest personalized behavior at item granularity. 
In this paper, we present the case for a coarse grained model which can capture the customer behavior at item category level.

The proposed Personalized Category (PC) model combined with Items-within-Category (IC) model outperform existing BIA and NBR models on standard public datasets. The PCIC model also scales well for large retailers with millions sized product catalogs and millions of active guests. The A/B tests on the site show a significant improvement in guest shopping experience and guest spends by using the model. 

In the future, we would recommend that retailers explore models that combine the insights from Personalized Category features with Personalized Item features. Moreover, we would recommend considering mutual excitation among items and categories as simultaneous consumption has some inherent relationship with repeat consumption.

\bibliographystyle{ACM-Reference-Format}
\bibliography{main}

\end{document}